\documentclass[twocolumn,aps,showpacs,pre,amsmath]{revtex4}
\usepackage{amssymb}
\usepackage{graphicx}

\begin{document}

\title{Disorder effects on energy bandgap and electronic transport \\ in graphene-nanomesh-based structures}

\author{V. Hung Nguyen$^{1,2}$\footnote{E-mail: viethung.nguyen@cea.fr}, M. Chung Nguyen$^2$, H. Viet Nguyen$^2$, and P. Dollfus$^3$}
\address{$^1$L$_-$Sim, SP2M, UMR-E CEA/UJF-Grenoble 1, INAC, 38054 Grenoble, France \\ $^2$Center for Computational Physics, Institute of Physics, VAST, P.O. Box 429 Bo Ho, Hanoi 10000, Vietnam \\ $^3$Institut d'Electronique Fondamentale, UMR8622, CNRS, Universite Paris Sud, 91405 Orsay, France}

\begin{abstract}
 Using atomistic quantum simulation based on a tight binding model, we investigate the formation of energy gap $E_g$ of graphene nanomesh (GNM) lattices and the transport characteristics of GNM-based electronic devices (single potential barrier structure and p-n junction) taking into account the atomic edge disorder of holes. We find that the sensitivity of $E_g$ to the lattice symmetry (i.e., the lattice orientation and the hole shape) is significantly suppressed in the presence of the disorder. In the case of strong disorder, the dependence of $E_g$ on the neck width is fitted well with the scaling rule observed in experiments [Liang \textit{et al.}, Nano Lett. \textbf{10}, 2454 (2010)]. Considering the transport characteristics of GNM-based structures, we demonstrate that the use of finite GNM sections in the devices can efficiently improve their electrical performance (i.e., high ON/OFF current ratio, good current saturation and negative differential conductance behaviors). Additionally, if the length of GNM sections is appropriately limited, the detrimental effects of disorder on transport can be avoided to a large extent. Our study provides a good explanation of the available experimental data on GNM energy gap and should be helpful for further investigations of GNM-based devices.

\end{abstract}

\maketitle

\section{Introduction}

In the recent years, thanks to its specific band structure and excellent carrier transport properties \cite{cast09,geim07}, graphene has been expected to be an excellent candidate for electronic applications \cite{schw10,mlin10,redd11}. However, the lack of energy gap between valence and conduction bands is a serious drawback regarding the operation of graphene transistors. In particular, the zero bandgap makes it very difficult to have a high ON/OFF current ratio and a really saturated current at high drain voltage \cite{meri08}. Fortunately, a number of techniques to open a bandgap in graphene have been proposed, for instance, cutting a graphene sheet into nanoribbons \cite{yhan07}, applying an electric field perpendicular to a bilayer graphene sheet \cite{oost08}, interaction of graphene with the substrate \cite{giov07,khar11}, or introducing doping and impurities in graphene \cite{hliu11,esca11}. This has hence stimulated a lot of investigations of various graphene nanodevices \cite{lian07,fior07,lian08,fior09,zhao09a,zhao09b,hung10,mich10,hung11,hung12b}.

Besides the finite bandgap graphene-based structures mentioned above, experimentalists have recently reported the fabrication of a new graphene nanostructure \cite{jbai10,mkim10,sini10,lian10} called graphene nanomesh (GNM) in which the size of nanoholes and the neck width (shortest distance between neighbor nanoholes) can be controlled down to the sub-10 nm scale. Various techniques have been developed to produce and control such GNM lattices, e.g., block copolymer lithography \cite{jbai10,mkim10}, nanosphere lithography \cite{sini10}, and nanoimprint lithography \cite{lian10}. This nanostructuring can open up a finite bandgap in large graphene sheets and hence GNM-based FETs can exhibit a similar ON/OFF current ratio as in graphene nanoribbon (GNR) devices \cite{jbai10}. Especially, GNM (2D semiconducting film) devices are able to carry electric current about 100 times greater than individual GNRs. The interest of GNMs is additionally due to the possibilities of bandgap engineering when varying the neck width \cite{jbai10,lian10}. The GNM lattices are hence expected to enlarge the applications of graphene-based devices, not only in electronics.

In this context, the theoretical analysis of electronic structure and transport properties of GNM lattices are very timely and desirable to provide a guidance for further investigations. Actually, a number of works \cite{pede08,wliu09,zhan11,ouya11,sahi11,oswa12,jipp11,guns11,kara11,yana12,hung12} have been carried out in this direction. The studies in \cite{zhan11,pede08,wliu09,ouya11,sahi11,oswa12,jipp11} have focussed on the electronic structure of perfect GNM lattices, especially, on the property of bandgap depending on the lattice constant, the hole shape and the orientation (i.e., holes along zigzag or armchair directions). Some of these works have tried to propose scaling rules to explain and predict the behavior of bandgap when varying the lattice parameters. There are also a few studies in which the transport properties of GNM-based structures have been investigated \cite{jipp11,guns11,kara11,yana12,hung12}. In particular, the transport properties related to the formation of GNM electronic structure were analyzed in \cite{jipp11}, the thermoelectric transport was considered in \cite{guns11,kara11,yana12}, while the design of negative differential conductance GNM-based devices was discussed in \cite{hung12}. However, to our knowledge, one important issue, the effects of disordered holes in GNM lattices, has not been studied yet. Therefore, our aim in this work is to clarify this point and we focus on two subjects: (1) the effects of edge disordered nanoholes on the energy gap of GNM lattices and (2) the optimization of the operation of some devices based on GNMs taking into account the disorder. The rest of paper is organized as follows. In section II, we introduce briefly the model and calculation methods which consist basically in using the non-equilibrium Green's function (NEGF) technique to solve a tight-binding Hamiltonian as in \cite{hung12}. The formation of the energy gap of GNM lattices is discussed in section III.A and the transport characteristics of single barrier structures and p-n junctions based on GNM lattices are presented in section III.B. Finally, the conclusion is drawn in section IV.

\section{Model and calculation}

In graphene nanostructures, the electronic properties of charges around the intrinsic (midgap) Fermi level can be described well by a single orbital tight binding approach \cite{pede08,wliu09,zhan11}. The corresponding Hamiltonian reads
\begin{equation}
{H_{tb}} = \sum\limits_n {{U_n}C_n^\dag {C_n}}  - t \sum\limits_{\left\langle {n,m} \right\rangle} {\left( {C_n^\dag {C_m} + h.c.} \right)}
\end{equation}
where $U_n$ is the on-site energy/potential energy at the \emph{n}-th site and $t = 2.7$ eV \cite{zhan11} denotes the hoping energy between nearest neighbor carbon atoms. This Hamiltonian has been recently used to investigate the electronic structure and the transport characteristics of perfect GNM-based devices \cite{hung12}. In such 2D structures, the extension along the transverse (OY) direction is considered through Bloch boundary conditions. Using the same approach as in \cite{hung12}, the considered lattice is split into elementary cells, the operators in equation (1) are then Fourier transformed along the OY direction and the Hamiltonian is rewritten in terms of decoupled (quasi-1D) forms
\begin{eqnarray}
{H_{tb}} &=& \sum\limits_{{k_y}} {{{\tilde H}_{1D}}\left( {{k_y}} \right)} \\
{{\tilde H}_{1D}}\left( {{k_y}} \right) &=& \sum\limits_n {\tilde H_n^ - \left( {{k_y}} \right) + {{\tilde H}_n}\left( {{k_y}} \right) + \tilde H_n^ + \left( {{k_y}} \right)} \nonumber
\end{eqnarray}
where ${{\tilde H}_n}\left( {{k_y}} \right)$ is the Hamiltonian of cell $\{ n \}$ and $\tilde H_n^\pm \left( {{k_y}} \right)$ denotes the coupling of cell $\{ n \}$ to cell $\{ n \pm 1\}$ along the transport direction OX.

Using the Hamiltonian (2), the device retarded Green's function in the ballistic approximation is computed for a given momentum $k_y$ as
\begin{equation}
\mathcal{G}\left( {\epsilon,{k_y}} \right) = {\left[ {\epsilon + i{0^ + } - {{\tilde H}_{1D}}\left( {{k_y}} \right) - \Sigma \left( {{k_y}} \right)} \right]^{ - 1}}
\end{equation}
where $\Sigma \left( {{k_y}} \right) = {\Sigma _L}\left( {{k_y}} \right) + {\Sigma _R}\left( {{k_y}} \right)$ with the self-energies $\Sigma_{L,R}$ describing the left and right contact-to-device coupling, respectively. The local density of states and the transmission probability needed to evaluate the current are calculated as $D(E,k_y,\vec{r}_n) = - {\rm{Im}}\left\{\mathcal{G}_{n,n}(E,k_y)\right\}/\pi$ and $\mathcal{T}\left( {\epsilon,{k_y}} \right) = {\rm{Tr}}\left\{ {{\Gamma _L}\mathcal{G}{\Gamma _R}{\mathcal{G}^\dag }} \right\}$, respectively, where ${\Gamma _{L\left( R \right)}} = i\left( {{\Sigma _{L\left( R \right)}} - \Sigma _{L\left( R \right)}^\dag } \right)$ is the transfer rate at the left (right) contact. The current is then computed using the Landauer formula
\begin{equation}
J = \frac{e}{{\pi h}}\int\limits_{BZ} {d{k_y}\int\limits_{ - \infty }^\infty  {d\epsilon\mathcal{T}\left( {\epsilon,{k_y}} \right)\left[ {{f_L}\left( \epsilon \right) - {f_R}\left( \epsilon \right)} \right]} }
\end{equation}
where ${f_{L\left( R \right)}}\left( \epsilon \right) = {\left[ {1 + \exp \left( {\left( {\epsilon - {E_{FL\left( R \right)}}} \right)/{k_b}T} \right)} \right]^{ - 1}}$ is the left (right) Fermi distribution function with the Fermi level $E_{FL\left( R \right)}$. At zero bias, the transport properties of the system is characterized by its conductance
\begin{equation}
G = \frac{{{e^2}L_G}}{{\pi h}}\int\limits_{BZ} {d{k_y}\int\limits_{ - \infty }^\infty  {d\epsilon\mathcal{T}\left( {\epsilon,{k_y}} \right)\left( { - \frac{{\partial f}}{{\partial \epsilon}}} \right)}}
\end{equation}
where the transverse width of graphene sheet $L_G$ is assumed to be much larger than the device length along the transport direction \cite{hung12}. The integrations over $k_y$ are performed in the first Brillouin zone.

The electronic structure and the transport in some GNM-based devices have been studied in \cite{pede08,wliu09,zhan11,ouya11,sahi11,oswa12,jipp11,guns11,kara11,yana12,hung12} but most of nanohole lattices are perfectly periodic. In practice, the disorder, e.g., the hole shapes may be randomly different, is inevitable in the fabrication process of GNM lattices \cite{jbai10,lian10}. Therefore, using the formalism above, we investigate the effects of this disorder on the electronic structure and the transport characteristics of GNM structures. To model the disorder, the carbon atoms in the edge of nanoholes are randomly removed with a uniform probability $P_D$ \cite{quer08,eval08,mucc09}. We focus on two different GNM lattices that consist of holes along the armchair (armchair holes) \cite{pede08} and zigzag (zigzag holes) \cite{zhan11} directions, respectively.

\section{Results and discussion}

\subsection{Energy gap of GNM lattices}

\begin{figure*}[!t]
\centering
\includegraphics[width=5.5in]{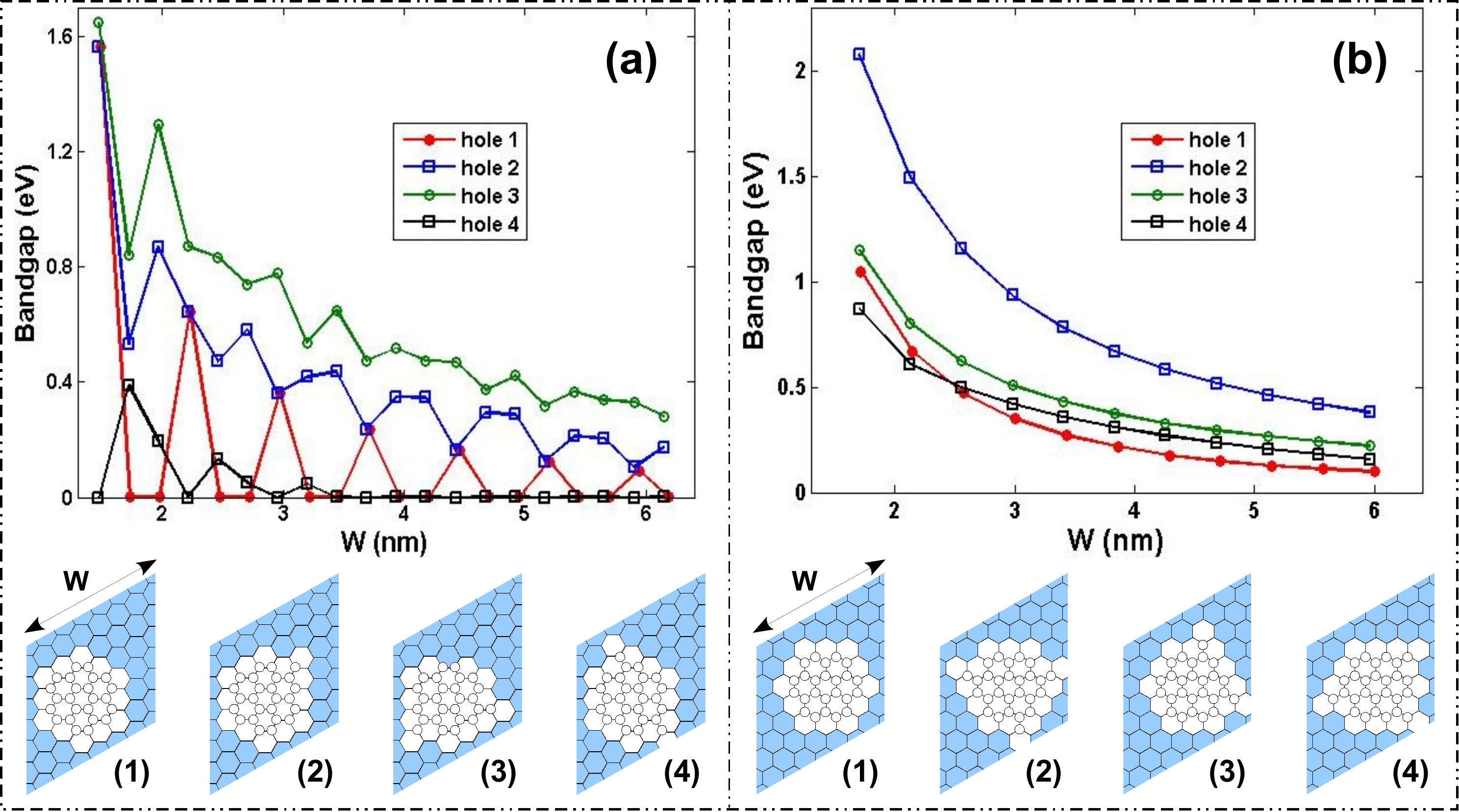}
\caption{Bandgap of GNM lattices as a function of the super-cell lattice constant for different hole shapes. (a) is for zigzag holes while (b) is for armchair holes. The super-cells of considered GNM lattices, where white circles indicate the removed carbon atoms, are presented in the bottom of two sub-figures.}
\label{fig_sim1}
\end{figure*}
First, we would like to review the recent studies on the electronic properties of GNM lattices. Strong efforts have been made on theoretical models and calculations on GNMs \cite{zhan11,pede08,wliu09,jipp11,ouya11,sahi11,oswa12} to predict the properties of their electronic structure and, especially, the possible scaling rule for bandgap $E_g$ with respect the shape, size, and architecture of holes introduced into the graphene sheet. However, the obtained results are very different. In particular, Pedersen \textit{et al.} \cite{pede08} considered the lattices of armchair circular holes and found a scaling rule $E_g \propto \sqrt{N_{rem}}/N_{tot}$, where $N_{rem}$ is the number of removed carbon atoms in a super-cell originally containing $N_{tot}$ atoms. However, this simple rule is not applicable for large $\sqrt{N_{rem}}/N_{tot}$ when $E_g$ fluctuates very strongly with the change in such the ratio. Moreover, together with the results obtained in \cite{wliu09}, it is shown that the behavior of $E_g$ is also dependent on the hole shape (circular holes in \cite{pede08}, triangular and rhombus holes in \cite{wliu09}). More significantly, the studies on the lattices of zigzag holes have shown very different results. Ouyang \textit{et al.} \cite{ouya11} predicted half of GNMs are semimetals and the rest are semiconductors, while it was shown in other works that only one-third of considered GNMs had a significantly finite bandgap \cite{zhan11,sahi11,oswa12} or even all GNMs are semiconductors \cite{wliu09}. The authors in ref. \cite{zhan11} also reported the scaling rule $E_g \propto 2^{-Q/3}$ for this kind of GNM lattice, where \emph{Q} is the super-cell lattice constant in the unit of $a_c = 2.46$ {\AA}. The complex properties of $E_g$ can be explained by the strong sensitivity of the GNM electronic structure to the hole shape and obviously to the lattice orientation, i.e., to whether we have zigzag or armchair holes.

Experimentally, Liang \emph{et al.} \cite{lian10} reported the bandgap scaling rule $E_g \propto 1/W_{nw}$ ($W_{nw}$ is the neck width), which can be explained by the important role on the GNM bandgap of the quantum confinement in the multiple graphene nanoribbon (GNR) network. A recent theoretical work \cite{jipp11} reported the same scaling rule for $E_g$ when studying the lattices of irregular holes but with a quite large distribution of $E_g$ at each $W_{nw}$. The latter feature indicates that the GNM bandgap is very sensitive to even a small change in the hole edges. It is important to note that neither metallic GNMs nor the orientation effects have been experimentally observed \cite{jbai10,mkim10,sini10,lian10}. As discussed above, the edge disorder of holes in the GNM lattice is inevitable \cite{jbai10,lian10}. Suggested by the studies \cite{quer08,eval08,mucc09} on disordered GNR structures, the edge disorder of holes may be an important factor to explain the discrepancies between theoretical works introduced above and experiments. To examine this idea, we model the disordered GNM lattices as follows. We assume that when fabricating GNM lattices with a given hole shape, a few percents of additional atoms at the edge of holes are unintentionally pulled out of the lattice. To model this system, we start with a (ordered/original) lattice of perfectly periodic holes and then randomly remove the edge atoms with a uniform probability $P_D$. This new (disordered) lattice has a mixing of a number of original holes and a few percents of holes having different shapes.
\begin{figure*}[!t]
\centering
\includegraphics[width=5.5in]{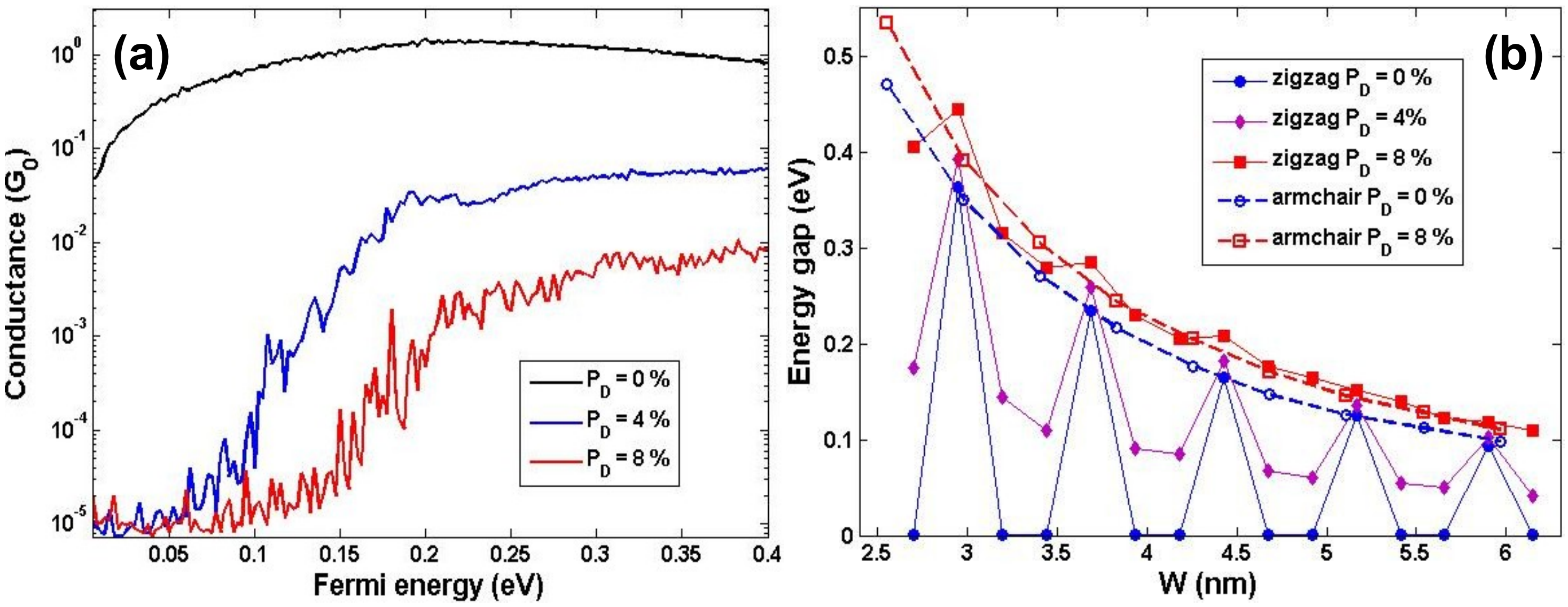}
\caption{(a) Conductance in the unit of $G_0 = e^2L_G/hW$ as a function of Fermi energy for different disorder strengths . The plot is for the lattices of zigzag holes with the hole shape 1 as shown in Fig. 1(a) and \emph{Q} = 13 corresponding to the metallic GNMs. (b) The energy gap of the GNM lattices as a function of the super-cell lattice constant with disorder effects. The plot is for the lattices of both zigzag and armchair holes with the hole shape 1 as shown in Fig.1(a) and (b), respectively.}
\label{fig_sim2}
\end{figure*}

We now use the tight binding Hamiltonian (1) to re-compute the bandgap $E_g$ of lattices of perfectly periodic holes to understand its dependence on the lattice parameters. In Fig.1, we plot $E_g$ as a function of the super-cell lattice constant \emph{W} (i.e., $W \equiv Qa_c$) for different hole shapes and for different orientations (zigzag holes in Fig.1(a) and armchair holes in Fig.1(b)). Generally speaking, it is shown that $E_g$ of semiconducting GNMs inversely decreases with an increase of the super-cell lattice constant $W$. Qualitatively, this is in good agreement with the experiments. However, it confirms again a big difference between the lattices of zigzag and armchair holes. In the case of zigzag holes, the results for the hole shape (1) (already studied in \cite{zhan11}) shows that a finite bandgap is observed when the index \emph{Q} is a multiple of 3 while all other GNMs are semimetallic. This feature has been explained in \cite{zhan11} as a consequence of the inter-valley scattering between different Dirac points of pristine graphene when the holes are created. However, our results in Fig.1(a) demonstrate this is not the case of other hole shapes even with a small change compared to the hole (1) and, simultaneously, the scaling rule $E_g \propto 2^{-Q/3}$ \cite{zhan11} is not applicable. For instance, the lattices with the hole shape (4) behave contrary to the previously suggested law, i.e, they are semimetals when \emph{Q} is a multiple of 3 while others are semiconductors. In this case, the GNM bandgap decays very rapidly with an increase of $W$. Besides, no semi-metallic GNMs with holes (2) and (3) are observed. Considering the lattices of armchair holes, we find from the results displayed in Fig. 1(b) that all studied GNMs are semiconductors. However, $E_g$ is also sensitive to the change in the hole shape. This sensitivity of $E_g$ to the hole shape is in good agreement with that reported in \cite{pede08,wliu09} since the ratio $\sqrt{N_{rem}}/N_{tot}$ of the considered lattices is large, i.e, $\geq 0.02$. Consistently with the previous works, our obtained results demonstrate again two important points for the lattices of perfectly periodic holes: (1) the GNM bandgap is very sensitive to the change in the hole shape and the lattice orientation; (2) it is hence difficult to determine theoretically a unique scaling rule to describe the bandgap of all different perfect GNM lattices.

Next, we go to investigate the disorder effects on the energy gap of GNM lattices. Here, we choose to present the results obtained for the lattices wherein the holes numbered (1) in Fig. 1, which are close to the holes fabricated experimentally \cite{jbai10,lian10}, are used as original holes. As explained above, the other holes (e.g., see in Fig. 1) probably occur in the considered lattices with a few percent fraction. As done in \cite{eval08,mucc09}, we use eq. (5) to compute the conductance at zero temperature and then measure the transport gap around the zero energy point, which implies the formation of the bandgap of the considered lattices \cite{lian10}. Note that all transport quantities are computed averagely over forty disorder samples. In Fig. 2(a), we plot the conductance of the GNM lattices as a function of energy and with different disorder strengths. In this figure, the lattices of zigzag holes with \emph{Q} = 13 are used. Similarly to the metallic GNRs studied in \cite{quer08,eval08,mucc09}, the ordered GNM lattice is a semimetal, i.e., no conduction gap is observed for $P_D = 0$ in Fig. 2(a). When the disorder occurs, the conductance is strongly suppressed, especially near the neutrality (zero energy) point. Close to the neutrality point, therefore, a deep conduction gap develops when increasing the disorder strength. The same feature is obtained for the lattices, which originate from the semiconducting lattices of perfectly periodic holes, but the enhancement of conduction gap is weaker than that in the metallic ones.

In Fig. 2(b), we display the energy (conduction) gap as a function of the super-lattice constant \emph{W} for different orientations (zigzag and armchair holes) and different disorder strengths. In what follows, we use the concepts "semiconducting (metallic) GNMs" to identify the lattices originating from semiconducting (metallic) GNMs of perfectly periodic holes. As discussed above, it is shown that while the conduction gap of semiconducting GNMs is slightly enlarged, the gap of metallic ones quickly develops with an increase of the disorder strength. Therefore, in the case of strong disorder, no metallic GNMs of zigzag holes are detected and the dependence of energy gap on \emph{W} tends to a unique scaling rule. Additionally, the results presented in Fig. 2(b) shows that the orientation effects are suppressed and the same scaling rule is obtained for two different orientations in disordered lattices, i.e., when $P_D = 8 \%$. More important, the energy gap in this case is fitted very well with the scaling rule $E_g = \alpha/W_{nw}$ \cite{lian10} with $\alpha \approx 0.6$ \emph{eVnm}. The difference between the value of $\alpha$ obtained here and $\alpha = 0.8\div0.95$ $eVnm$ in \cite{lian10} can be explained by the fact that our considered holes are smaller and the disorder seems to be weaker than in the case studied in their work. Besides, our simulations show that even in the presence of weak disorder, the orientation effects are always suppressed and the scaling rule $E_g = \alpha/W_{nw}$ is a good approximation for different original holes (not displayed here), though $E_g$ (or $\alpha$) is still sensitive to a significant change in their shape. However, this sensitivity of $E_g$ to the hole shape is also suppressed in the presence of strong disorder (very large $P_D$) when all different holes contribute equivalently to the electronic properties of the sample.

These results are very similar to those observed for GNRs \cite{quer08,eval08,mucc09,yhan07} and demonstrate that the edge disorder of holes is one of important factors which weaken the lattice symmetry effects and make the effects of quantum-confinement in the nanoribbon crossing network of GNM lattices dominant. It explains well the behavior of energy gap observed in experiments \cite{lian10}.
\begin{figure*}[!t]
\centering
\includegraphics[width=5.5in]{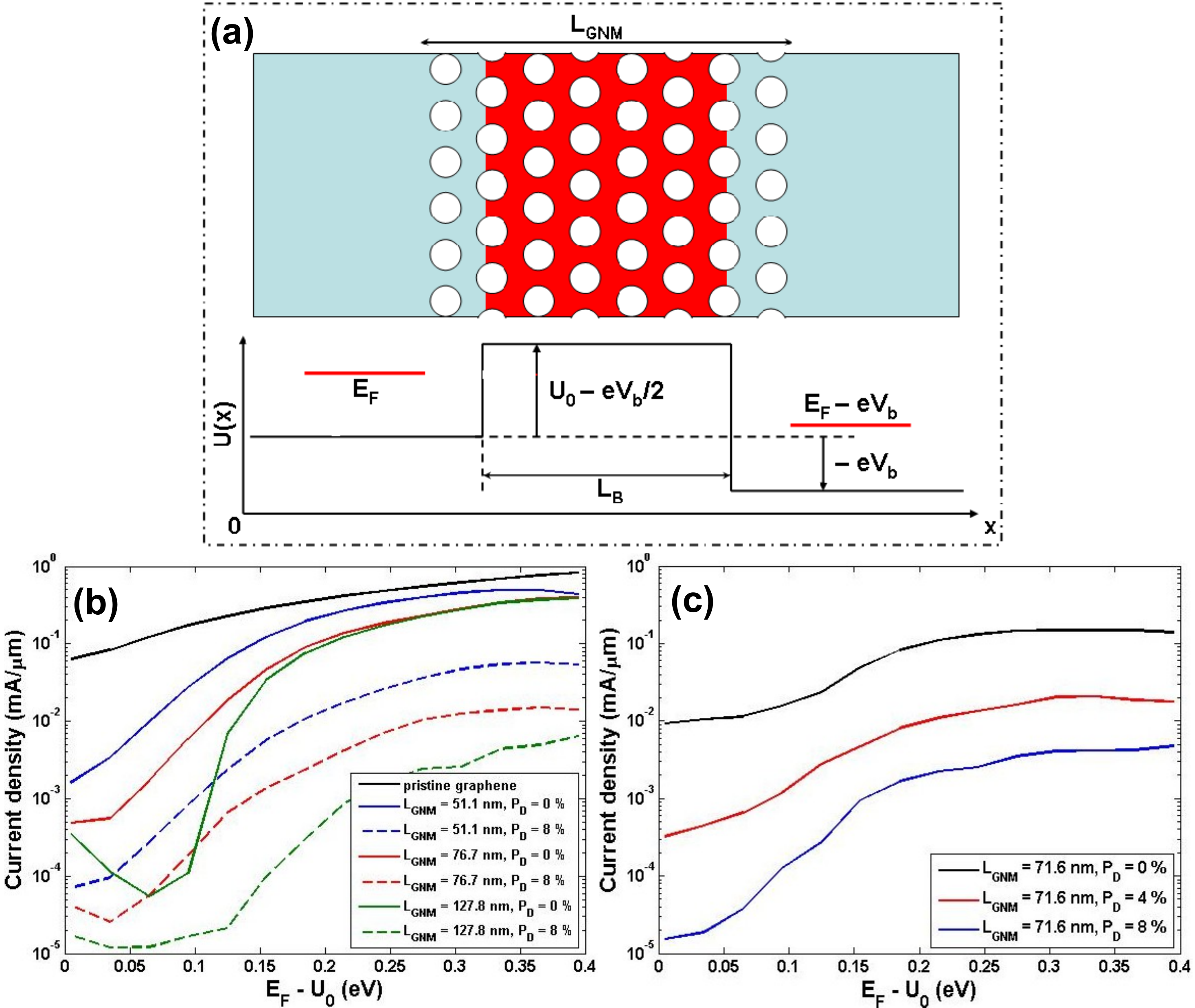}
\caption{(a) Schematic of potential barrier structures studied in this work. A finite GNM section is inserted in the device center. (b) and (c) show the current density at $V_b = 25$ mV as a function of the barrier height with the barrier length $L_B \thickapprox 61.8$ nm and $E_F = 0.4$ eV. In (b), the results are plotted for \emph{Q} = 15 ($W \thickapprox 3.7$ nm, semiconducting GNM), for different $L_{GNM}$ and disorder strengths. (c) is for \emph{Q} = 14 ($W \thickapprox 3.45$ nm, metallic GNM), and for different disorder strengths. In (b) the case of full pristine graphene structure is shown for comparison.}
\label{fig_sim3}
\end{figure*}

\subsection{Transport characteristics of GNM-based structures}

In this sub-section, we discuss the use of the GNM lattices to optimize the operation of graphene electronic devices including the disorder effects. We focus on two typical devices based on a single (gate-induced) potential barrier structure and a p-n junction. In what follows, the zigzag holes are used and all simulations were performed at room temperature.

\subsubsection{Single potential barrier structures}

The schematic of the simulated single-barrier structure is shown in Fig. 3(a), wherein the potential barrier can be generated and controlled by a gate electrode as in field-effect-transistors \cite{alar12}. The model used here is not self-consistently coupled to Poisson's equation. Though ideal and simple, it can provide the basic and most important information on the transport behavior and the electrical characteristics of devices \cite{vnam08} such as the graphene FET. One of the important properties of transistors is their ability to switch off the current by tuning the gate voltage, i.e., by changing the potential barrier. In principle, to significantly switch off the current, i.e., to obtain a high ON/OFF current ratio, a large bandgap of the channel material is essentially required. Therefore, the 2D monolayer graphene devices have a serious drawback inherent in the semi-metallic character of the conducting material. Using a GNM lattice with finite bandgap \cite{jbai10,lian10} offers a possibility to overcome this limitation more conveniently than by using very narrow GNRs \cite{yhan07} or bilayer graphene with strong vertical electric field \cite{oost08}.
\begin{figure*}[!t]
\centering
\includegraphics[width=5.5in]{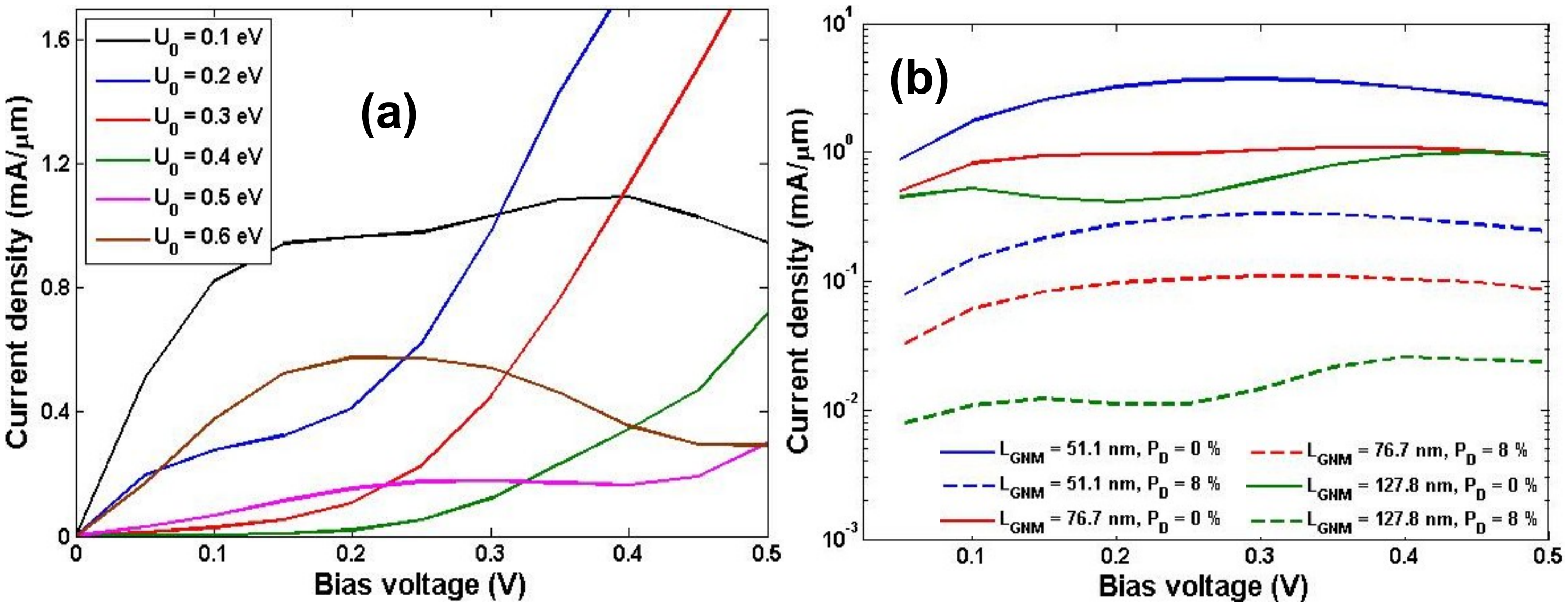}
\caption{(a) \emph{I-V} characteristics of the potential barrier structure with a finite perfect GNM section and for different barrier heights. (b) shows the effects of different $L_{GNM}$ and disorder strengths for the case of $U_0 = 0.1$ eV in (a). The device parameters are $E_F = 0.4$ eV, $L_B \thickapprox 61.8$ nm, $L_{GNM} \thickapprox 76.7$ nm, and \emph{Q} = 15.}
\label{fig_sim4}
\end{figure*}

Obviously, the disorder is not only shown to influence strongly the band structure, it is also expected to significantly affect the transport quantities such as the conductance and the current \cite{yoon07,basu08}. However, to obtain a low OFF current, a finite bandgap material is mandatory in the gated part of the channel, but not in two source and drain access regions (or leads here). Therefore, we propose to use the graphene device as schematized in Fig. 3(a) with a finite GNM section inserted in between pristine graphene leads to limit at the minimum the GNM-induced effects of disorder while the good controllability of the current can still be guaranteed. We plot in Fig. 3(b) the current density as a function of the barrier height $U_0$ for different lengths of GNM section and in two cases of ordered and disordered lattices. Here, the semiconducting GNM with \emph{Q} = 15 is used. It is clearly shown that the OFF current (ON/OFF ratio) is strongly reduced (enhanced) when inserting a finite GNM section into the device, i.e., compare the current for $L_{GNM} \neq 0$ with that for $L_{GNM} = 0$ (pristine graphene). Very significantly, in the presence of disorder, the effect of increasing $L_{GNM}$ is not only to reduce the OFF current and to enhance the ON/OFF ratio, but also to reduce the ON current. To have both high ON current and high ON/OFF ratio, our results provide the important suggestion that the optimum device should be determined by the competition between the effects of finite bandgap and disorder of the GNM section used. The best device parameter ($L_{GNM}$) is thus dependent on the bandgap value and on the disorder strength, i.e., to keep constant the ON (OFF) current, the shorter (larger) $L_{GNM}$ is required if the disorder (bandgap) is stronger (smaller). For instance, we suggest that $L_{GNM} \simeq 51 \div 77$ nm shall be a good choice for the devices studied in Fig. 3(a), i.e, the ON/OFF ratio is about a few hundreds while the ON current is still high.

For the devices with a metallic GNM section, the disorder not only results in the suppression of the current but also essentially originates the energy gap opening. To see the disorder effects in this case, we display in Fig. 3(b) the current density for different disorder strengths. If the holes are perfectly periodic, it is shown that the current is smaller than that in normal graphene devices (see in Fig. 3(a)). This is nothing but the consequence of an enhancement of the channel resistance when inserting the nanoholes in the device. However, due to the lack of finite bandgap, the ON/OFF current ratio is quite similar in both cases. Interestingly, when the disorder occurs in the GNM section, though the current is reduced, the ON/OFF ratio significantly increases, i.e., it is about 16, 65 and 310 for $P_D = $ 0 $\%$, 4 $\%$ and 8 $\%$, respectively. When comparing the two cases of semiconducting and metallic GNM devices, we observe that the current and the ON/OFF ratio tend to the same scale when the disorder is strong enough. This result is very consistent with the property of energy gap discussed above. Finally, we emphasize again that for both of these cases, the ON/OFF ratio may increase but the ON current decreases with an increase of $L_{GNM}$ in the disordered devices and hence a finite GNM section shall be mandatory to ensure good device operation.
\begin{figure*}[!t]
\centering
\includegraphics[width=5.0in]{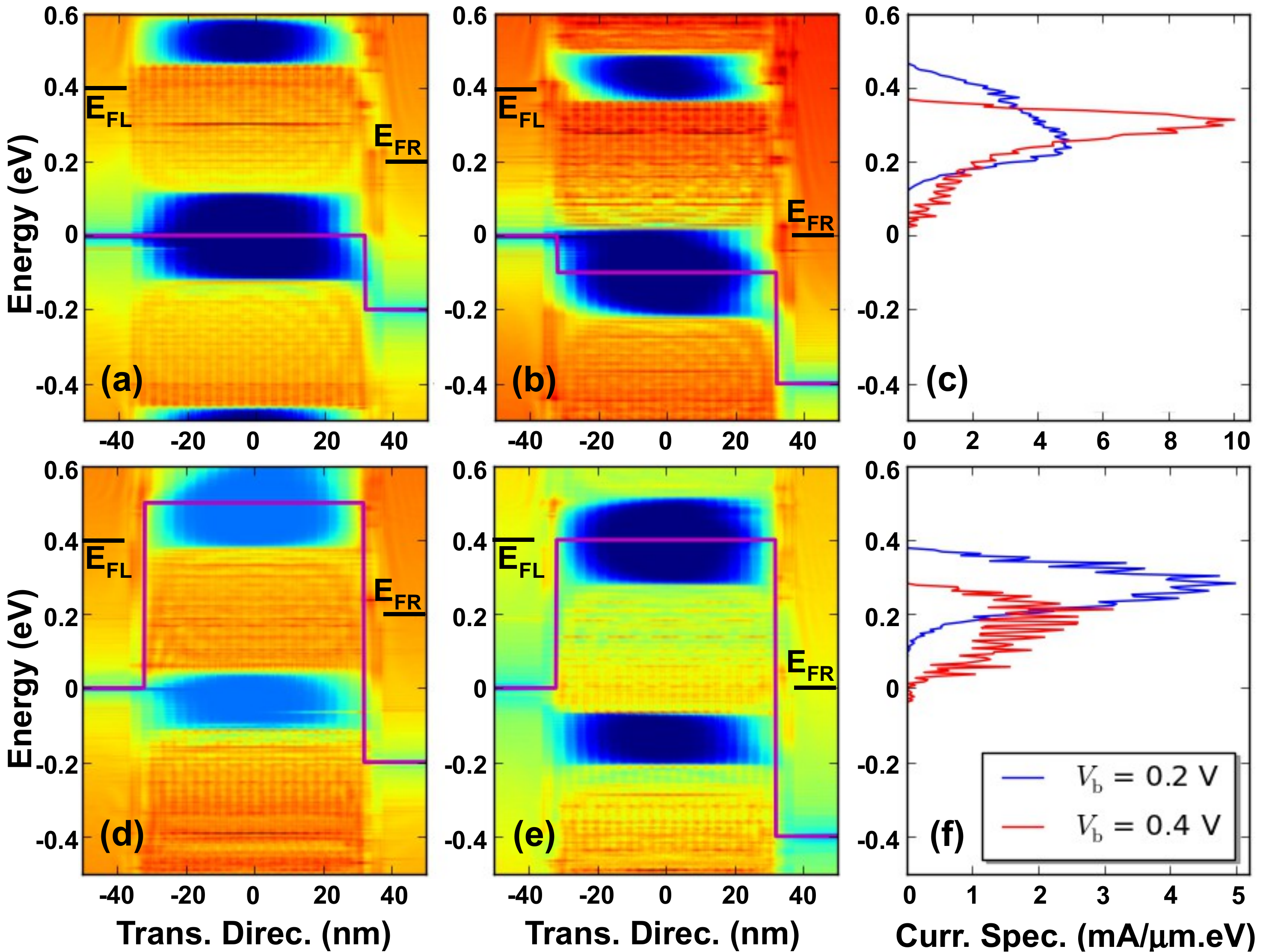}
\caption{(a,b,d,e) Local density of states and (c,f) current spectrum at two different biases ($V_b = 0.2$ and 0.4 V) for the device studied in Fig. 4(a). The barrier heights are $U_0$ = 0.1 eV for (a,b,c) and 0.6 eV for (d,e,f). The solid line in (a,b,d,e) indicates the potential profile.}
\label{fig_sim4}
\end{figure*}

Regarding the electronic applications, besides the improvement of the ON/OFF current ratio in the graphene devices, recent studies \cite{meri08,thie10,jbai11,yawu12,szaf12} have been done to achieve and control a low output conductance in saturation or even a negative differential conductance (NDC). This leads to the improvement of voltage gain, one of important figures of merit for analog high frequency applications. In gapless monolayer graphene devices the current saturation is relatively weak \cite{meri08,alar12} because of the important contribution of band-to-band tunneling \cite{vnam08}. In principle, a good current saturation behavior is observed only in transistors made of large bandgap channel materials such as conventional semiconductors, semiconducting carbon nanotubes, graphene nanoribbons and of course graphene nanomeshes. It is hence important to analyze the output conductance and the current saturation in our simulated GNM devices and to evaluate the impact of GNM disorder, though the GNM section is finite and limited to the active device region. In Fig. 4(a), we display the \emph{I-V} characteristics obtained for different barrier heights. Here, the semiconducting GNM lattice of perfectly periodic holes is used while the barrier length is fixed ($L_B \simeq 61.8$ nm). In the normal graphene devices, a weak current saturation can be observed and explained by the effects of the transmission valleys occurring around the Dirac (neutrality) points in the left contact and the channel \cite{alar12,vnam08}. When rising the bias voltage, while the extension of the energy window of current transmission $[E_{FR}, E_{FL}]$ tends to increase the current, the two transmission valleys mentioned above when approaching the window $[E_{FR}, E_{FL}]$ can reduce the output conductance. Hence, the output conductance is fully dependent on the way these transmission valleys reach the window $[E_{FR}, E_{FL}]$. In particular, the lowest output conductance (current saturation or even NDC) should be achieved when the Dirac point $E^d_C$ in the channel approaches $E_{FL}$ at the same bias as $E_{FR}$ reaches the Dirac point $E^d_L$ in the left contact. This suggests a way to improve such the behavior by using a finite bandgap graphene channel to make the transmission valleys wider and deeper \cite{alar12}. It is exactly what we propose in this work: using a finite GNM section in the barrier region to enlarge the transmission valley around $E^d_C$. Indeed, the results presented in Fig. 4(a) illustrate clearly this idea, i.e., a good current saturation (or even NDC) behavior is observed for some potential barriers.

In particular, the low output conductance can be achieved in two regimes: very low and high barriers. To explain its origin, we display in Fig. 5 the local density of states and the current spectrum as a function of energy at two different biases $V_b = 0.2$ V and 0.4 V and for two different barrier heights $U_0 = 0.1$ eV and 0.6 eV. For $U_0 = 0.1$ eV, when rising the bias voltage, $E^d_C$ moves down in energy to come close to and/or even below $E^d_L$. Hence, the transmission valley (wide and deep compared to that in normal graphene devices) around $E^d_C$ strongly limits the current spectrum close to $E_{FR}$ (probably including the band-to-band tunneling \cite{vnam08}) at high bias, i.e., see in Figs. 5(a,b,c). Moreover, besides the main energy gap around the neutrality point, some mini-gaps emerge at high energy in the band structure of GNMs \cite{hung12}. Here, the first mini-gap in the GNM conduction band generates an additional transmission valley, which can reduce the current component close to $E_{FL}$ for $U_0 = 0.1$ eV when rising $V_b$. Based on these features, the current (with a low output conductance) at high bias in this case is smaller than for $U_0 = 0.2$ or 0.3 eV. For high barriers (e.g., $U_0 = 0.5$ and 0.6 eV), the transmission valley around $E^d_C$ reach the energy level $E_{FL}$ at high bias, which reduces strongly the current component close to $E_{FL}$, while the transmission valleys around $E^d_L$ and due to the first mini-gap in the GNM valence band limit the current close to $E_{FR}$, i.e., see in Figs. 5(d,e,f). The output conductance is hence small and can be negative when rising the bias voltage. For instance, for the barrier $U_0 = 0.6$ eV $\equiv 3/2 E_F$, $E^d_C$ approaches $E_{FL}$ at the same bias $V_b = 0.4$ V as $E_{FR}$ reaches $E^d_L$ and a strong NDC behavior is hence achieved.

To clarify the disorder effects, we plot in Fig. 4(b) the \emph{I-V} characteristics for different $L_{GNM}$ and disorder strengths while the potential barrier is fixed ($U_0 = 0.1$ eV). First, considering the case of perfectly periodic holes, the current is shown to decrease when increasing $L_{GNM}$. This feature is observed because the finite bandgap material (GNM) is used not only in the barrier region but also in the two contacts when $L_{GNM} > L_{B}$. Hence, when increasing $L_{GNM}$ ($> L_B$), the effects of energy gap in the GNM sections in the two contacts can affect and reduce the current, which is consistent with the results presented in Fig. 3(a). More important, the results for disordered devices in Fig. 4(b) demonstrate again that though the low output conductance is always observed at high bias, an increase of $L_{GNM}$ enhances the disorder effects, i.e., reduces significantly the current.

All results presented in this sub-section suggest that graphene devices can be optimized by inserting only a (appropriately) finite GNM section into the channel to avoid the disorder effects while good characteristics such as high ON/OFF current ratio and strong current saturation (or even NDC behavior) are still guaranteed. For instance, we propose to use $L_{GNM} \simeq 51 \div 77$ nm for the devices considered here.

\subsubsection{P-N junctions}

\begin{figure}[!t]
\centering
\includegraphics[width=3.4in]{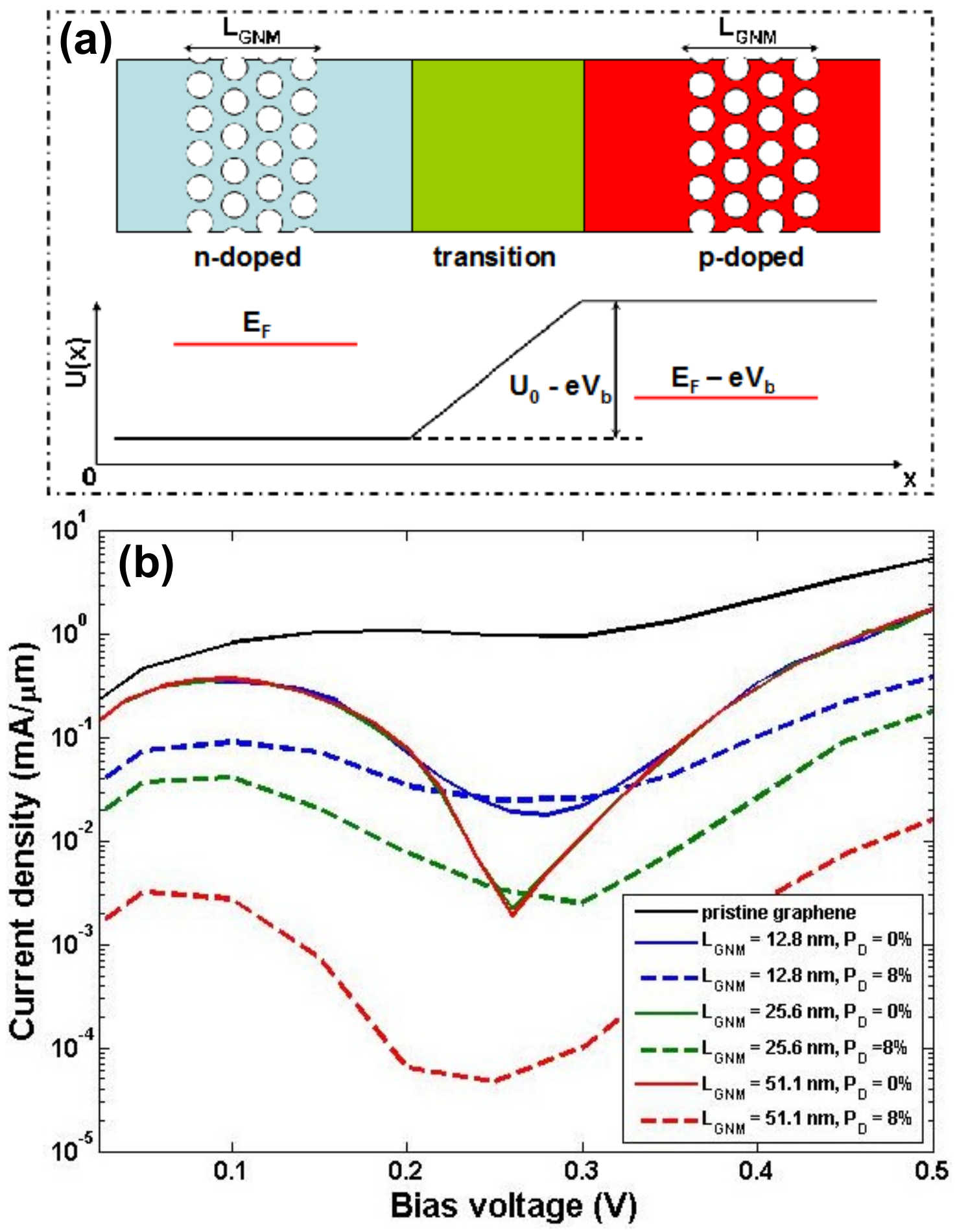}
\caption{(a) Schematic of simulated graphene p-n junctions with finite GNM sections inserted in the two doped contacts. (b) shows the obtained \emph{I-V} characteristics with the effects of different $L_{GNM}$ and disorder strengths. The zigzag holes with \emph{Q} = 15 are used while $E_F = 0$ and $U_0 = 0.5$ eV.}
\label{fig_sim5}
\end{figure}
Now, we discuss the use the GNM lattices to improve the operation of graphene $P-N$ junctions. The $P-N$ junctions are very well-known electronic structures, since they can exhibit a non-linear \emph{I-V} characteristics with a strong NDC behavior \cite{hung11,hung12}, which offers great potential for high frequency applications \cite{mizu95}. In principle, this NDC behavior occurs at high bias when the filled states in n-doped side find a reduced number of empty states available for tunneling in the p-doped side. The feature can be observed essentially due to the large bandgap materials used in two doped contacts, e.g., GNRs in \cite{hung11} and GNMs in \cite{hung12}. Besides, it was also shown that the current peak in this kind of device is strongly dependent on the bandgap and the length of the transition region in the device center. Therefore, we have proposed in \cite{hung11,hung12} to use graphene hetero-structures with a gapless (or small bangap) in the transition region to obtain high current peak and peak-to-valley current ratio (PVR). However, there is still one issue, the disorder effects, which has not been investigated yet in \cite{hung12} and will be clarified below.

As suggested by the transport features discussed above, it may not be necessary to use a very long (semi-infinite) section of semiconducting material in two doped contacts to obtain a strong NDC effect in the P-N junctions. A finite semiconducting GNM section is expected to be enough while limiting the influence of disorder effects. To examine this idea, we consider the P-N junctions schematized in Fig. 6(a) and plot their \emph{I-V} characteristics in Fig. 6(b) for different lengths of GNM sections and for both two cases of with and without disorder. The devices considered here are similar to that investigated sec. 3.3 of ref. \cite{hung12} where a normal graphene section is inserted in the transition region, but the very long (quasi-infinite) GNM sections  in the contacts are replaced by the finite ones. Indeed, without the disorder, it is shown that (i) when inserting finite GNM sections into the doped contacts, a strong NDC behavior (compared to the normal graphene p-n junction) is clearly observed; (ii) the PVR of NDC behavior increases and saturates when $L_{GNM}$ is long enough, e.g., $L_{GNM} > 26$ nm for the devices considered here. It is worth noting that a strong NDC behavior with PVR of $\sim 200$ is observed in Fig. 6. However, similarly to the features observed above, we find that the current is significantly affected by the edge disorder of holes, i.e., though a high PVR is still possible, the current peak is strongly reduced when increasing the GNM length. In particular, the PVR is about 4, 17 and 70 while the observed current peak is 92, 41, and 3.2 $\mu$A/$\mu$m for $L_{GNM} = 12.8$, 25.6 and 51.1 nm, respectively. On this basis, to obtain a high PVR and, simultaneously, to avoid the strong suppression of the current due to the disorder, it suggests to use (appropriately) finite GNM sections in two doped contacts. For instance, $L_{GNM} \simeq 25 \div 52$ nm may be a good choice for the devices simulated in Fig. 6.

\section{Conclusion}

In this work, using atomistic quantum simulation within a tight-binding model solved by the NEGF technique, we have investigated the properties of energy gap of GNM lattices and considered the optimum graphene devices for electronic applications using finite GNM sections including the edge disorder of holes. We found that the dependence of the GNM energy gap $E_g$ on the lattice symmetry (i.e., the lattice orientation and the hole shape) is significantly suppressed and the quantum-confinement effects in the nanoribbon crossing network of GNMs become dominant in the presence of the disorder. When a strong disorder occurs, the dependence of $E_g$ on the neck width is fitted well with the scaling rule ($E_g = \alpha/W_{nw}$) observed in experiments \cite{lian10}, which is an important result. Regarding the operation of graphene devices, our study focusses on single potential barrier structures and p-n junctions. It demonstrates that to obtain good performance (i.e., high ON/OFF current ratio, strong current saturation and NDC behaviors), the use of finite GNM sections in the devices can be a very good option. More important, since they are unavoidable, the proposed configurations are shown to be able to avoid the strong disorder effects which tends to limit the output current through the devices. Though more accurate calculation models (e.g., self consistent simulations with solving the Poisson equation) are needed for a more quantitative evaluation of device performance, our work provides a good guidance for further investigations of GNM-based devices.

\section*{Acknowledgment}

One of the authors (P.D.) acknowledges the French ANR for financial support under the projects NANOSIM-GRAPHENE (ANR-09-NANO-016) and MIGRAQUEL (ANR-10-BLAN-0304). The work in Hanoi was supported by the Vietnam's National Foundation for Science and Technology Development (NAFOSTED).

\end{document}